# DOES FULLERIDE SUPERCONDUCTIVITY "CROSSOVER" CONTAIN A FESHBACH RESONANCE?

Richard H. Squire [ξ], Norman H. March [*†]

[ξ] Department of Chemistry
West Virginia University – Institute of Technology
Montgomery WV 25136

[*] Oxford University, Oxford, England
[†] Department of Physics
University of Antwerp
Antwerp, Belgium

Abstract

The pioneering works by Eagles, Leggett, and Nozieres/Schmitt-Rink [1, 2, 3] (reviewed and augmented by Randeria [4]) emphasize that in the limits of the models studied both at $T = 0$ and $T \neq 0$, the "cross-over" from a BCS-type to a BEC-type superconductor is continuous. The BCS and BEC "end points" seem to be well-established. However, in the intermediate region – home to fulleride and high temperature superconductors – considerable extrapolation of the models must be done as there still is no exact theory. Yet, considerable current literature is devoted to what appears to be more "singular-type phenomena" such as quantum critical points, "stripe" formation, insulator to superconductor phase transitions, loss of validity of the Fermi liquid theory, etc.

Using a connection we have made with "cold" atom Fermion-Boson crossover theory [5], we can establish that the resonance previously discussed [6] is a result of the crossing of the Fermion band by the Boson band. While the ground and singly excited states appear to remain continuous, the collective modes due to the resonance transform the nature of the superconductivity. We discuss features of the resonance and the experimentally observed pre-formed "BEC Cooper pair" formation in fullerides, essential to the Boson-Fermion resonance theory. In addition some of the various singular phenomena discussed above can be put into perspective.

1. Introduction.

The study of BCS to BEC crossover has been studied in an almost continuous manner since the discovery of high temperature superconductors (high-$T_c$). In many of the studies the legacy of the pioneering work of Eagles [1], Leggett [2], and Nozieres and Schmitt-Rink (NSR [3]) remains. A recent additional flux of interest was injected with the demonstration that atomic systems containing cold Fermionic atoms can become bosons with the assistance of a Feshbach resonance [4]. The resonance seems to imply a discontinuity between the Fermion ("BCS-like) side and the Boson (BEC-like side), as both the BCS and BEC extremes are anchored by well established, exact theories [5].

Indeed, considerable recent literature is devoted to what appears to be "singular-type phenomena" such as quantum critical points, insulator to superconductor phase transitions, loss of validity of the Fermi liquid theory, etc. [6]. This manuscript discusses some of these apparent contradictions using a model system of fulleride superconductor (SC) we have developed.

In our previous work on fulleride SC [7] we identified the two electron-doped fulleride state as an insulator with Wigner-like correlations. Here we extend this characterization and suggest that the pairing of the two doped electrons occurs via many-body Cooper pairing interaction with the energy lowering being large enough to permit pairing, thereby contradicting Hund's rule for the $t_{1u}$ Jahn-Teller level. The pairing further opens a gap in the singlet particle spectrum leading to a "pseudogap" and a modified phase diagram. We previously discussed a "soft" electron-hole pair associated with the very broad three electron $t_{1u}$ plasmon band at 0.5 eV (band width 0.5 eV) [8]. This plasmon seemed to resemble a resonance, possibly similar to the Anderson - Kondo model and we recognized, as Ranninger had pointed out [9], that the BFM contains a similar effect. Further pressing the analogy of the BFM to a Feshbach resonance in recent cold atom work (Ohashi and Griffin [11]) leads to a more complete explanation of fulleride superconductivity as an example of BCS to almost BEC crossover (section 5).

What follows in section 2 is a background of the crossover to put certain concepts in place such as the limiting cases of BEC and BCS theory. Section 3 summarizes a version of the Boson-Fermion Model (BFM) which is widely discussed in "cold" Fermion atom research. We then directly compare the fulleride model rather favorably to the "cold" atom results in Section 3. We finish with a summary, conclusions, and our proposal for future work.

2. Perspective of BEC - BCS Crossover

To provide continuity with previous crossover studies (and the associated assumptions) in a dilute gas [12], a functional integral formulation is outlined of a three dimensional continuum model that will allow us to put our model in perspective. The Hamiltonian is

$$H = \bar{\psi}_\sigma(x)\left[-\frac{\nabla^2}{2m} - \mu\right]\psi_\sigma(x) - g\bar{\psi}_\uparrow(x)\bar{\psi}_\downarrow(x)\psi_\downarrow(x)\psi_\uparrow(x)$$

where the chemical potential $\mu$ fixes the average density, g is the strength of the coupling, the system is in a unit volume, and $\hbar = k_B = 1$. At a temperature $\beta^{-1}$ the partition function Z is written as an imaginary time functional integral with action (with $x = (\bar{x}, \tau)$)

$$S = \int_0^\beta d\tau \int d\bar{x}\, \{\bar{\psi}_\sigma(x)\partial_\tau \psi_\sigma(x) + H\}$$

The fermion interaction can be decoupled with the Hubbard-Stratonovich field $\Delta(\bar{x}, \tau)$, then the fermions can be integrated out obtaining [13]

$$Z = \int D\Delta D\bar{\Delta}\, \exp{-S_{eff}[\Delta, \bar{\Delta}]}$$

$S_{eff}$, the effective action is

$$S_{eff} = \frac{1}{g}\int_0^\beta d\tau \int d\bar{x}\, |\Delta(x)|^2 - Tr \ln G^{-1}[\Delta(x)]$$

and $G^{-1}$ is the inverse Nambu propagator

$$G^{-1}(x, x') = \begin{pmatrix} -\partial_\tau + \frac{\nabla^2}{2m} + \mu & \Delta(x) \\ \Delta^*(x) & -\partial_\tau - \frac{\nabla^2}{2m} - \mu \end{pmatrix}$$

The saddle-point approximation ($\Delta \equiv 0$) is stable at high T; it will become unstable below a certain temperature, $T_0$. The transition temperature is defined by

$$\delta S_{eff}/\delta\Delta[\Delta = 0] = 0$$

and the solution can be written as

$$\frac{1}{g} = \sum_k \tanh\left(\frac{E_k}{2T_0}\right)/2E_k \qquad (1a)$$

where $E_{\bar{k}} = \sqrt{\xi_{\bar{k}}^2 + \Delta_0^2}$, $\xi_{\bar{k}} = \varepsilon_{\bar{k}} - \mu$ and $\varepsilon_{\bar{k}} = \bar{k}^2/2m$. The ultraviolet divergence in the above equation is regulated by replacing the bare coupling constant g by the low energy limit of the two-body t-matrix resulting in

$$m/4\pi a_s = -1/g + \sum_{|\bar{k}|<\Lambda}(2\varepsilon_k)^{-1} \qquad (1a')$$

which defines the s-wave scattering length, $a_s$. The bare interaction $1/a_s$ increases monotonically from $-\infty$ for a weak interaction to $+\infty$ for a strong attraction [5]. Beyond the two-body bound state threshold $1/a_s = 0$, $a_s$ is the size of the bound state with binding energy $E_b = 1/ma_s^2$. The dimensionless coupling constant is then $1/k_F a_s$ which covers the complete BCS-BEC range, varying from $-\infty$ in the weak coupling BCS to $+\infty$ in the BEC limit [14].

Using eq. (1a) the equation for the transition temperature in terms of the renormalized coupling $a_s$ and "crossover parameter" $(k_F a_s)^{-1}$ is

$$-\frac{C_0}{k_F a_s} = \sum_{\bar{k}} \left[ \frac{\tanh\left(E_{\bar{k}}/2T_0\right)}{2E_{\bar{k}}} - \frac{1}{2\varepsilon_{\bar{k}}} \right]$$

where $C_0 = mk_F/4\pi$. As there are two unknowns, $T_0$ and $\mu$, we need a second equation (as shown by BCS) to fix the number of particles, the so-called "number" equation, $N = -\partial\Omega/\partial\mu$ as a Lagrange multiplier; stated differently, this fixes the chemical potential for a given density. The saddle point approximation leads to the expression for the thermodynamic potential $\Omega_0 = S_{eff}[\Delta_0]/\beta$ which, when simplified, results in

$$N = \sum_{\bar{k}} \left[ 1 - \frac{\xi_{\bar{k}}}{E_{\bar{k}}} \tanh\left(E_{\bar{k}}/2T_0\right) \right] \quad (1b)$$

This is the standard BCS solution when $\mu = \varepsilon_F$; however in the crossover $\mu$ is a strong function of the coupling in the BEC region where all of the particles are affected by the attraction. These self-consistency conditions are standardly given by two BCS-like equations referred to as the "gap" (eq 1a) and "number" (eq 1b) equations, respectively. The thermodynamics of condensed and broken pairs are accounted for in eq (1b), but not the collective modes due to pair excitations.

Several additional calculations are discussed in the Randeria review article [5], including Gaussian fluctuations about a broken symmetry saddle-point that represents collective excitations of the superconductor. In the BCS limit these involve broken pairs that are pushed to high energy with increasing attraction. In the Bose limit the collective modes are the dominant low energy excitations. The conclusion is that the intermediate coupling range needs to include both of these types of excitations. Incorporating these terms numerically shows the spectrum smoothly interpolates between the Bogoliubov sound mode in strong coupling and the Anderson mode in weak régime. The eigenvector for q = 0 is a pure phase Goldstone mode with mixing controlled by deviation from particle-hole symmetry. In a charged system there are two points of interest for the fulleride discussion later: 1) at low density a Wigner crystal should be expected (or at least a solid with Wigner crystal-like correlations as we have surmised for fullerides [15]), and 2) the collective modes discussed above get modified into plasmons since the system is charged. Finally, there is a schematic phase diagram (Fig 1a) that illustrates a crossover below which pair correlations and dissociation temperature are important along with the transition temperatures for a lattice and continuum model. In the broken symmetry state $(<T_c)$ as attractions increase, the system evolves from a BCS state with large, overlapping Cooper pairs to a Bose condensate of tightly bound composite bosons and the Fermi surface disappears. A modified fulleride phase diagram is presented for comparison (Fig 1b); note the presence of a pre-formed pair (see sections 3 and 4 for details).

3. "Cold" Fermion Crossover Using the Boson-Fermion Model (BFM).

The BFM allows an enhancement of the BCS to BEC crossover described above by incorporating a new, strongly attractive interaction between Fermions mediated by the quasi-molecular Boson associated with a Feshbach resonance [4, 7, 11]. Then, applying BCS theory to a degenerate Fermi gas with a strong pairing interaction results in a strong suppression of $T_{BCS}$ caused by fluctuations in the two-particle Cooper channel. In the strong coupling regime there are two "types" of Bosons even above $T_c$, "molecules" ("molecules" in "cold" atom parlance) associated with the resonance and pre-formed Cooper pairs, as previously established [3, 4], see Fig 1b. A crossover from BCS to BEC will show dramatic changes due to the resonance.

To make a connection with the atomic work on cold superfluid Fermi gases [10, 11, 16] we rewrite our previous Hamiltonian (following Ohashi and Griffin, and Chen et al) and keep only the essential terms for the non-resonant and resonant Cooper pair / Boson molecule interaction (see [7], Appendix C for the full Hamiltonian):

$$H = \sum_{\bar{p},\sigma} \varepsilon_{\bar{p}} c^{\dagger}_{\bar{p}\sigma} c_{\bar{p}\sigma} - U \sum_{\bar{p},\bar{p}'} c^{\dagger}_{\bar{p}\uparrow} c^{\dagger}_{-\bar{p}\downarrow} c_{-\bar{p}'\downarrow} c_{\bar{p}'\uparrow} \quad (2)$$

$$+ \sum_q \left(E^0_{\bar{q}} + 2\nu\right) b^{\dagger}_{\bar{q}} b_{\bar{q}} + g_r \sum_{\bar{p},\bar{q}} [b^{\dagger}_{\bar{q}} c_{-\bar{p}+\bar{q}/2\downarrow} c_{\bar{p}+\bar{q}/2\uparrow} + h.c.]$$

Here $c_{\bar{p}\sigma}$ and $b_{\bar{q}}$ represent the annihilation operators of a Fermion (Fermi atom) with kinetic energy $\varepsilon_{\bar{p}} = p^2/2m$ and a quasi-molecular Boson with the energy spectrum $E^0_{\bar{q}} + 2\nu = q^2/2M + 2\nu$, respectively. In the second term $-U < 0$ is the BCS theory attractive interaction from non-resonant processes while the threshold energy of the composite Bose particle energy band is denoted by $2\nu$ in the third term. The last term is the Feshbach resonance (coupling constant $g_r$) that describes how a b-Boson (again, a "molecule" in "cold" atom parlance) can dissociate into two Fermions, or how two Fermions can bind into a b-Boson. Since the b-Boson "molecule" is constructed from a bound state consisting of two Fermions, the boson mass is $M = 2m$ and the conservation of total number of particles N imposes a different number relationship than previously

$$N = \sum_{\bar{p}\sigma} \left\langle c^{\dagger}_{\bar{p}\sigma} c_{\bar{p}\sigma} \right\rangle + 2 \sum_{\bar{q}} \left\langle b^{\dagger}_{\bar{q}} b_{\bar{q}} \right\rangle \equiv N_F + N_B \quad (3)$$

Incorporating this constraint into eq (2) again results in a grand canonical Hamiltonian, as used previously for variable particle number, since b bosons ("molecules") are formed from fermions, and vice-versa. With this relationship, there is only one chemical potential,

$$H - \mu N = H - \mu N_F - 2\mu N_B$$

and it leads to an energy shift

$$\varepsilon_{\vec{p}} \to \xi_{\vec{p}} \equiv \varepsilon_{\vec{p}} - \mu$$

and

$$\varepsilon_{B\vec{q}} + 2\nu \to \xi_{B\vec{q}} \equiv \varepsilon_{B\vec{q}} + 2\nu - 2\mu$$

From this point we outline a particle-particle vertex and mean-field solutions [13] that provide most of the essential features necessary for the fulleride superconductor phase diagram. We emphasize that these BFM solutions were derived for cold fermion work, but the reformulation exposes what appears to be a surprisingly close connection between the cold Fermion work and fulleride SC.

Following [11] a superfluid phase transition temperature can be determined by the Thouless criterion that describes the instability of the normal phase of Fermions due to an attractive interaction allowing formation of Cooper pairs. Calculating a four-point vertex function provides an equation for the particle-particle vertex with a solution

$$\Gamma(\vec{q}, i\nu_n) = -\frac{U - g^2 D_0(\vec{q}, i\nu_n)}{1 - [U - g^2 D_0(\vec{q}, i\nu_n)]\Pi(\vec{q}, i\nu_n)} \quad (4)$$

where $D_0^{-1}(\vec{q}, i\nu_n) = i\nu_n - E_{\vec{q}}^0 - 2\nu + 2\mu$ is the correlation function of the Cooper pair operator of two Fermions (atoms in cold atom parlance) with total momentum $\vec{q}$ defined by $\hat{\Delta}(\vec{q}) \equiv \sum_{\vec{p}} c^\dagger_{\vec{p}+\vec{q}/2,\uparrow} c^\dagger_{-\vec{p}+\vec{q}/2,\downarrow}$ in the absence of U and $g_r$. The $\Pi(\vec{q}, i\nu_n)$ term is the particle-particle propagator describing Cooper pair fluctuations, needed to describe a non-BCS state in the crossover (not considered in section 2),

$$\Pi(\vec{q}, i\nu_n) = \sum_{\vec{p}} \frac{1 - f(\varepsilon_{\vec{p}+\vec{q}/2} - \mu) - f(\varepsilon_{\vec{p}-\vec{q}/2} - \mu)}{\varepsilon_{\vec{p}+\vec{q}/2} + \varepsilon_{\vec{p}-\vec{q}/2} - 2\mu - i\nu_n} \quad (5)$$

and $f(\varepsilon)$ is the Fermi distribution. When the particle-particle vertex (eq. 4) develops a pole at $\vec{q} = \nu_n = 0$, a superfluid phase transition occurs which corresponds to the following "gap" equation for $T_c$, to be contrasted with eq. (1a),

$$1 = \left(U + g_r^2 \frac{1}{2\nu - 2\mu}\right) \sum_{\vec{p}} \frac{\tanh(\varepsilon_{\vec{p}} - \mu)/2T_c}{2\varepsilon_{\vec{p}} - 2\mu} \quad (6)$$

In eq (6) $g_r^2/(2\nu - 2\mu)$ is the additional pairing interaction mediated by a boson which becomes very large when $2\mu \to 2\nu$. $T_c$ is the temperature at which an instability occurs in the normal phase of a degenerate Fermi gas due to formation of bound states with zero center of mass momentum $(\vec{q}=0)$ and energy $2\mu$. A connection can be made with cold atom theory by rewriting eq (1a') as

$$\frac{m}{4\pi a_s} = \frac{1}{U_{eff}} + \sum_{\vec{k}} \frac{1}{2\varepsilon_{\vec{k}}} \quad (6a)$$

with $U \to U_{eff} = U + \frac{g_r^2}{2\mu - 2\nu}$ and invoking the "fundamental postulate of crossover theory" that the superconductivity varies smoothly through the Feshbach effect [4].

The chemical potential for this model is determined from eq. (3), ($\mu = \varepsilon_F$ in BCS theory) assuming $\mu$ is temperature independent. However, $\mu$ has a more fundamental deviation when the quasi-molecules with $\vec{q} \neq 0$, pre-formed Cooper pairs, and superfluid fluctuations are all present. The "number" equation for Fermions (atoms) $N(\mu,T)$ can be obtained from the thermodynamic potential $\Omega$ as stated previously, $N = -\partial\Omega/\partial\mu$. NSR studied the fluctuations previously [3]; what is new here is the term originating from the Feshbach coupling of the b-Bosons and Fermions in eq. (5). Then,

$$N = N_F^0 + 2N_B^0 - T\sum_{\bar{q}} e^{i\delta\nu_n} \frac{\partial}{\partial\mu} \ln\left[1 - \left(U - g_r^2 D_0(q)\right)\Pi(q)\right] \quad (7)$$

where $N_F^0 \equiv 2\sum_{\vec{p}} f(\varepsilon_{\vec{p}} - \mu)$ and $N_B^0 = \sum_{\bar{q}} n_B(E_{\bar{q}}^0 + 2\nu - 2\mu)$ with $n_B(E)$ the Bose distribution function. Again, the "gap" and "number" equations (eq (6) and (7)) must be solved self-consistently. An intuitive interpretation can be obtained by use of the identity $N_B^0 = -T\sum_{\bar{q},\nu_n} D_0(\vec{q},i\nu_n)$, and rewriting eq. (7) as

$$N = N_F^0 - 2T\sum_{\bar{q}} \tilde{D}(q) - T\sum_{\bar{q}} \frac{\partial}{\partial\mu_F} \ln\left[1 - U_{eff}(q)\right]_{\mu_F \to \mu} \quad (8)$$

$$\equiv N_F^0 + 2N_B + 2N_C$$

(The separation of the "two types" of two-particle bound states or Bosons $2N_B$ and $2N_C$ in eq (8) is an attempt to provide further insight.) The second term is a renormalized b-Boson Green function,

$$\frac{1}{\tilde{D}^{-1}(q)} = i\nu_n - E_{\bar{q}}^0 - 2\nu + 2\mu - \Sigma(q)$$

with the self-energy $\Sigma(q) \equiv -g_r^2 \tilde{\Pi}(q)$ and $\tilde{\Pi}(q) \equiv \Pi(q)/[1 - U\Pi(q)]$. This is interpreted as the b-Bosons contribution as affected by the self-energy. The third term $2N_C$ is similar to the fluctuations studied by NSR but now including the Cooper pair fluctuations with the effective interaction $U_{eff}(q) = U - g_r^2 D_0(q)$ which now depends on energy as well as momentum.

Transforming the Matsurbara frequency summation into a frequency integration for the renormalized b-Bosons ($N_B$ in eq.(8)), one finds

$$N_B = -\frac{1}{\pi}\sum_{\vec{q}} \int_{-\infty}^{+\infty} dz\, n_B(z)\, \text{Im}\,\tilde{D}(\vec{q}, iv_n \to z + i\delta)$$

Taking the principle value in the z-integration, we find **that if the b-Boson (molecule) decays into two Fermions in the presence of the Feshbach resonance, it has a finite lifetime** given by the inverse of the imaginary part of the self-energy $-g_r^2 \tilde{\Pi}$ in $\tilde{D}(\vec{q}, z - i\delta)$. But **when the chemical potential becomes negative by lowering the threshold energy $2v$, the renormalized b-Bosons do not decay if their energies are smaller than $E_{\vec{q}}^0 - 2\mu$** since it can be shown from eq. (5) that $im\Pi(\vec{q}, z + i\delta)$ is proportional to the step function $\Theta(z + 2\mu - E_{\vec{q}}^0)$. The energy of a stable molecule $(\omega_{\vec{q}})$ corresponding to the pole of $\tilde{D}$ is given by

$$\omega_{\vec{q}} = (E_{\vec{q}}^0 - 2\mu) + [2v - g_r^2 \tilde{\Pi}(\vec{q}, \omega_{\vec{q}})] \quad (9)$$

if $E_{\vec{q}}^0 - 2\mu > 0$. Long-lived Bosons (molecules) appear when the renormalized threshold $2\tilde{v} \simeq 2v - g_r^2 \tilde{\Pi}(\vec{q}, z)$ becomes negative as decay into two Fermions is forbidden. $N_B$ in eq.(11) consists of two kinds of Bosons, stable ones $N_B^{\gamma=0}$ with infinite lifetime, and quasi-Bosons $N_B^{\gamma>0}$ which can decay into two Fermions. The contribution of the poles describing the stable Bosons gives

$$N_B^{\gamma=0} = \sum_{\vec{q}}^{poles} Z(\vec{q}) n_B(\omega_{\vec{q}}) \quad \text{with} \quad Z(\vec{q})^{-1} = 1 + g_r^2 \frac{\partial \tilde{\Pi}(\vec{q}, \omega_{\vec{q}})}{\partial z}$$

describing the mass renormalization. The third term in eq.(8) describing Cooper pair fluctuations can be analyzed similarly. The non-resonant s-wave interaction U can be rewritten in terms of the s-wave scattering length $U = 4\pi a N / m$, then $U/\varepsilon_F = 0.85(p_F a)$. For a dilute Fermi gas $p_F a \ll 1$, so $U/\varepsilon_F \ll 1$. Then, the non-resonant attractive interaction U cannot generate pre-formed Cooper pairs. **But, when $2\mu$ approaches $2v$, the interaction $U_{eff}(q)$ mediated by the b-Bosons (molecules) becomes so strong that the pre-formed Cooper pairs appear, as suggested in [3] (see Fig 1).** The energy of these poles is the same as given in eq. (9). This enables us to "divide" $N_C$ into contributions from stable pre-formed Cooper pairs $(\equiv N_C^{\gamma=0})$ and scattering states $(\equiv N_C^{SC})$. If $g = 0$ or $\tilde{\Pi}$ is ignored, there are no pre-formed Cooper pairs and $Z(\vec{q}) = 1$ in

$$N_C^{\gamma=0} = \frac{g_r^2}{2} \sum_{\vec{q}}^{poles} \frac{\partial \tilde{\Pi}(\vec{q}, \omega_{\vec{q}})}{\partial \mu} Z(\vec{q}) n_B(\omega_{\vec{q}})$$

The model Hamiltonian gives a rather clear picture of the BCS to BEC crossover. When the threshold $2v$ is larger than $\varepsilon_F$, the Fermi states are dominant. A BCS-like phase transition is found since $\mu \sim \varepsilon_F > 0$ and stable bosons are absent since $\text{Im}\,\tilde{\Pi}(\vec{q}, z) \neq 0$ for $z > 0$. The only exception is for $z = 0$, then $\text{Im}\,\tilde{\Pi}(\vec{q},0) = 0$. If $2\tilde{\nu} = 2\mu$ so a stable Boson with $\vec{q} = 0$ appears at $\omega_{\vec{q}} = 0$ eq. (9), this condition reduces to the gap equation for $T_c$, eq. (6). This will result in the formation of stable Cooper pair Bosons and a phase transition at the same temperature, namely the BCS theory. In this limit no stable, long-lived Bosons with $\vec{q} \neq 0$ exist above this transition temperature.

In the $\nu \ll 0$ limit, the Fermi states are almost empty and we expect the phase transition to be BEC-like. Stable Bosons can appear even above $T_c$. The phase transition of these stable Bosons occurs when the energy of the Boson $(\vec{q} = 0)$ reaches zero as measured from the chemical potential. Again eq. (4) results and the problem is a BEC transition of a non-interacting gas of $N/2$ Bosons with mass $M = 2m$ and no free Fermions. Eq. (4) gives $2\mu = 2\tilde{\nu}$ and a Bose condensate appears in $N_B^{\gamma=0}$ and $N_B^{\gamma \neq 0}$.

The fulleride SC is between the two limits in the previous two paragraphs. It is interesting to note that eq. (6) and (1a), derived as a condition for a superfluid phase transition via formation of two particle bound states, also describes BEC in a gas of stable b-Bosons and pre-formed Cooper pairs.

4. Fulleride Superconductivity as an Example of A Resonant Crossover.

The reformulation of the BF Model in "cold atom" terms made the connections with fulleride SC more apparent, recognizing, of course, that fulleride SC is not a limiting case of BEC. The result of doping of two electrons into a fulleride crystal produces a Jahn-Teller (JT) effect which results in a orbital triplet, $t_{1u}$. The two doped electrons localized on a fulleride molecule in a singlet, violating Hund's rule and the result is an insulating state. There are several suggestions as to how this might occur [17, 18, 19]. We have an alternative suggestion, namely that the spin triplet state is unstable with respect to the formation of a localized "BEC Cooper pair" (CP) above $T_c$ and below some $T^*$ such that the energy gained (significantly stronger than a BCS CP) is sufficient to negate Hund's rule. (This CP could be considered as the elusive "pre-formed pair" heretofore speculated about but not experimentally identified in high-$T_c$ studies.) The result is that the JT orbital triplet is split with the pair in the occupied orbital now lower in energy by an amount $E_I$. Now, the pseuodogap origin seems apparent; it is due to a localized pairing without long range order (LRO), see Fig 1b, as the phases of the wave functions of adjacent fullerides are random. This proposal also suggests why the metal-insulator (M-I) transition that certain doped fullerides span happens; small changes in a fulleride structure can move seemingly closely related fullerides from one side of the transition to the other. The possibility that doping a third electron onto a fulleride can cause an energy change larger that energy $E_I$ would suggest that the metal would have three unpaired electrons. This possibility is inconsistent with our superconductivity theory, so we are left with the interesting possibility that the pre formed pair remains intact upon doping with a third electron. The resulting material might be perceived as a "poor conductor" as the possibility exists that the preformed pair remains localized. This material, say $Rb_3C_{60}^{3-}$, might be described as a "boson conductor." It is interesting to note that with a modest change of energies in Gunnarsson's arguments [19], one can arrive at our conclusion even though that was not the original intent; this is a reflection of the subtlety of this particular M-I transition where it seems that structurally related fullerides must be determined case by case as to whether they are metals or insulators.

Examining the possible structures for the fulleride pre-formed CP using the Weisskopf model [20] where one electron moves in one direction and creates a "tube" of attraction for the second electron moving in the opposite direction, a fulleride with two doped electrons is an insulator, i.e. spins are paired and angular momentum is zero. An important question is, "Are all of the $H_g$ symmetry vibrations involved in the CP formation, or are there only a few, energy-selected (resonant) modes?" If we assume only a few vibrational modes are involved, then the CP could be anisotropic and this might be a resolution of the debate over whether pre-formed CP's or vortices are the first aspect of high-$T_c$'s; this fulleride model suggests that the CP on fulleride might be viewed as either [4].

This argument receives experimental support from the interpretation of the Raman spectra obtained by Winter and Kuzmany [21]. The dramatic changes in the linewidths and level shifts of the two highest and two lowest $H_g$ vibrational modes suggest that the fulleride CP might have an anisotropy due to the strong resonance coupling. In addition the Rb atoms in $Rb_3C_{60}^{3-}$ were expected to have two inequivalent octahedral and tetrahedral positions in the Rb NMR. Experimentally the tetrahedral site is split which also could be explained by an anisotropic CP [22]. It seems that the standard BCS theory offers little hope of providing any explanation. Certainly if the CP suggestion is correct, then it seems likely that the spatial extent of a fulleride CP places it between a "real space" and a momentum space pair. This is not inconsistent with recent work on high-$T_c$ systems [23]. The end result, regardless of whether this simple model is correct or not, is that fullerides should possess a defined pre-formed CP, a necessary and seemingly illusive point in applying the BF Model to high-$T_c$ systems.

Another major difference in the "cold" Fermi-Boson treatment above and the fulleride superconductivity (SC) is that instead of "tuning" $\nu$ using a magnetic field, fulleride SC is tuned by the amount of doped electrons. Starting from the BCS side (Fig 1b with four doped electrons (n = 4), when the threshold $2\nu$ is much larger than $\varepsilon_F$, the Fermi states are dominant. A BCS-like phase transition is found since $\mu \sim \varepsilon_F > 0$ and stable bosons are absent ($\operatorname{Im}\tilde{\Pi}(\vec{q},z) \neq 0$ for $z > 0$). The only exception is for $z = 0$ ($\operatorname{Im}\tilde{\Pi}(\vec{q},0) = 0$). If $2\tilde{\nu} = 2\mu$ a stable Boson with $\vec{q} = 0$ appears at $\omega_{\vec{q}} = 0$ in eq. (9), and this condition reduces to the gap equation for $T_c$, eq. (6). This will result in the formation of stable Cooper pair Bosons and a phase transition at the same temperature, namely the BCS theory. In this limit no stable, long-lived Bosons with $\vec{q} \neq 0$ exist above this transition temperature. As doping decreases to n = 3 electrons, the apex in the phase diagram is approached which seems to be close to the crossover point $(\mu = 0)$, and molecules with a finite lifetime begin to be formed which is equivalent to $2\nu \leq 2\varepsilon_F$. Figure 2 illustrates the connections of the ground state free energy with the location of the Boson and Fermion bands and the resonance interaction for fullerides.

Finally, below doping levels of 2.5 electrons, superconductivity ceases as itinerant "third" electrons are needed to sustain pairing of b-type Bosons. As doping of 2 electrons is approached, a fulleride molecule maintains a 'pre-formed" Cooper pair (c-type Bosons) instead of following Hund's rule where the two electrons would possess parallel spins in the $t_u$ Jahn-Teller state (Fig 1b). The resulting state has Wigner crystal-like correlations and the ordered state is accompanied by the appearance of a pseudogap that first began forming when three electrons are doped on a fulleride [15].

As mentioned, fulleride SC is not in the BEC limit; if we were dealing with a **true BEC system, in the $\nu \ll 0$ limit**, the Fermi states would be almost empty and we would expect the phase transition to be BEC-like. The phase transition of these stable Bosons occurs when the energy of the Boson $(\vec{q} = 0)$ reaches zero as measured from the chemical potential. Again eq. (6) results and the problem is a BEC transition of a non-interacting gas of $N/2$ Bosons with mass $M = 2m$ and no free Fermions. Eq. (6) gives $2\mu = 2\tilde{\nu}$ and a Bose condensate appears in $N_B^{\gamma=0}$ and $N_B^{\gamma \neq 0}$. It is interesting to note that eq. (6) and (1a), derived as a condition for a superfluid phase transition via formation of two particle bound states, also describes BEC in a gas of stable b-Bosons and pre-formed Cooper pairs.

Is there a Quantum Critical Point (QCP) in the fulleride model? The gap to single particle excitations are given by the minimum of the Bogoliubov quasiparticle energy

$$E_{gap} \equiv \min_{\varepsilon_k \geq 0} \left[ (\varepsilon_k - \mu)^2 + |\Delta_k|^2 \right]^{1/2}$$

If $\mu > 0$, the minima occur at $\varepsilon_k = \mu$ (BCS) and the energy gap is the gap parameter $\Delta$. As the attraction increases, as some point the chemical potential will go below the bottom of the band where $\varepsilon_k = 0$ and $E_{gap} \neq \Delta$. The gaps to single particle excitation in the s-wave case for both conditions are

$$E_{gap} = \Delta \text{ For } \mu > 0$$

$$\text{and } E_{gap} = (\mu^2 + \Delta^2)^{1/2} \text{ for } \mu < 0$$

This indicates a weak singularity at $\mu = 0$. At $\mu = 0$ there seems to be a demarcation point between BCS $(\mu > 0)$ and BEC $(\mu < 0)$. As concluded by Engelbrecht et al [12], it does not appear that a singularity is present in the quasiparticle energy. There is no doubt that with a Feshbach resonance present, there is a singularity in the two body scattering length.

Even though the two body scattering length changes abruptly at the unitary scattering condition, superconductivity still varies smoothly. However, there is a fundamental change in the nature of the superconductivity and Cooper (or "molecular") pairs as the previously suggested location [15] of a critical point in the fulleride phase diagram is passed. In addition some features of the evidence for a QCP such as quasiparticle lifetime variation are present [24, 25, 26]. Using the connection we have established with the cold atom work, it seems obvious that at T = 0 the Feshbach resonates dominates. One of the features of a QCP is presumably its far reaching influence which is clearly present in the Feshbach resonance. Clearly more work needs to be done to better understand the dominance of the resonance interaction over the non-resonance interaction, and exactly what features might better defined a QCP [26, 27].

5. Summary, Conclusions, and Future Work.

Initially we established the Eagles, Leggett, NSR, Randeria context of the crossover from BCS to BEC with a gap and number equations, eq (1a) and (1b), respectively. This helped establish the continuity of the ground state free energy with a well defined value of the product $k_F \xi$

The recognition that a pre-formed pair is formed near the n = 3 doping level ($\mu \simeq 0$) entirely changes the nature of the crossover on the "under-doped" side of the SC "dome" (from n = 3 to n = 2). Experimental evidence for the pre-formed boson at n = 2 was presented. The BF Model applied to high-$T_c$ has a premise about a pre-formed boson, but less is well known about its microscopic origin. In some reports the hole doped in the copper oxides does not lead to metallic behavior, possibly similar to fulleride behavior, but much more work needs to be completed. Certainly the "pre-formed" boson – from the perspective of n = 2 singlet in the triply degenerate $(t_{1u})$ orbital lends a good deal of credence to the BF Model as applied to fullerides. As the doping is reduced from n = 4, and the concomitant free energy reduced, the boson band approaches the Fermion band until a resonance condition is established, presumably around the maximum $T_c$. We now have a possible explanation for cause of the resonance noted previously [7].

It is interesting to note the features of the BF Model in common with the Anderson [28] and Kondo models [29], and heavy electron theory [27, Table 1]. This fulleride SC model might well be a simple prototype for certain features of high-$T_c$ compounds. It seems likely that the work of Tesanovic et al [23] in establishing a micro-domain in high-$T_c$ suggests an arena where a pre-formed pair might exist, but clearly much more work needs to be done in this area.

As to future directions, further experimental work on the fullerides is called for. The Wigner-like correlations should be one area of focus. Certainly elucidating more explicitly the nature of the MI transition and the role of the pre-formed pair could offer significant insight, and might suggest other compounds where pre-formed pairs could exist., and enable a calculation of the condensation energy.

Lastly, we can suggest further study of the nature of the QCP and the strength of the effective interaction $(U_{eff})$ at resonance. Is this what a universal QCP might look like? Can Coleman's table [27] be quantitatively explained?

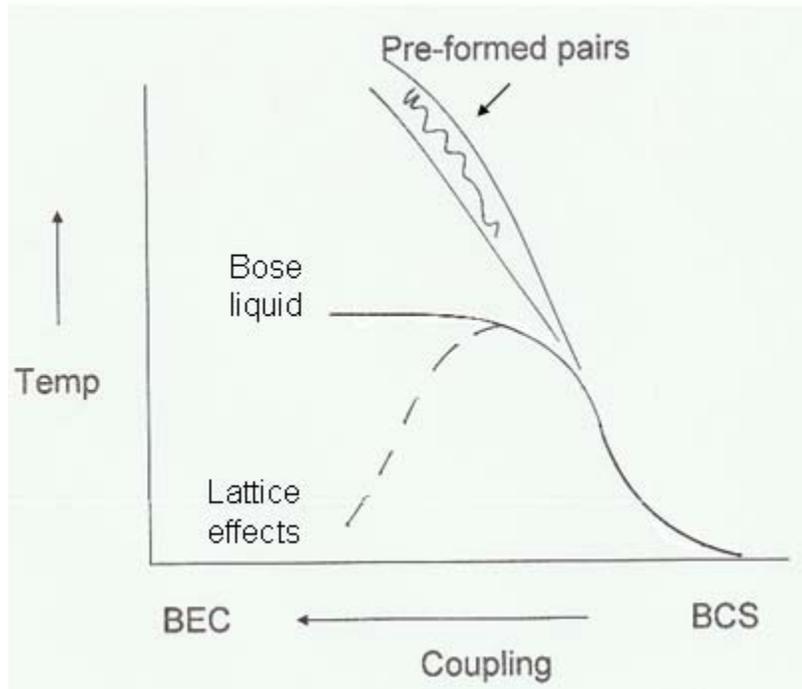

Fig 1a. Qualitative phase diagram for attractive Fermions. The pre-formed pair (broad segment) illustrates a crossover region, below which pre-formed pairs exist, as in the n = 2 electron doped fulleride (singlet). The full curve is the transition temperature for a continuum model, with the dashed line representing lattice effects. Note the similarities with the fulleride phase diagram in Fig 1b.

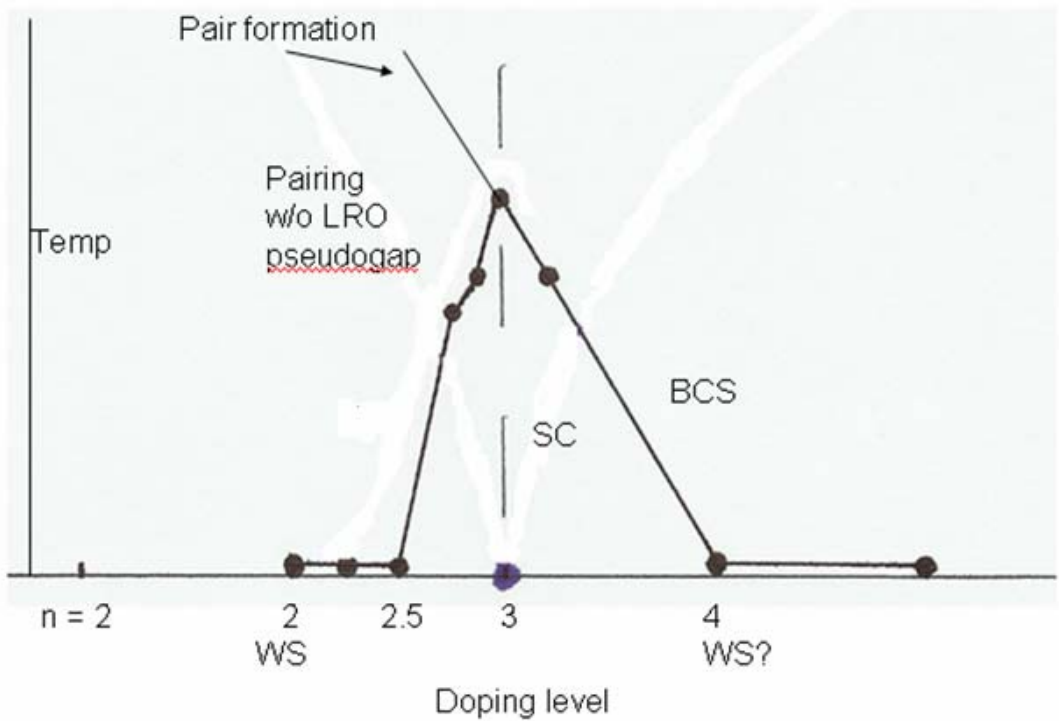

Fig 1b Revised fulleride phase diagram. Pre-formed pairs are the result of the Cooper pairing energy being large than the exchange energy for parallel spins in the Jahn-Teller triplet $t_{1u}$ state. The pair is also stabilized in the BEC region (see text). The underdoped region results from resonant Cooper pairing, while the BCS regime is the result of the non-resonant portion from the pairing term, $U_{eff} = U + \dfrac{g_r^2}{2\mu - 2\nu}$

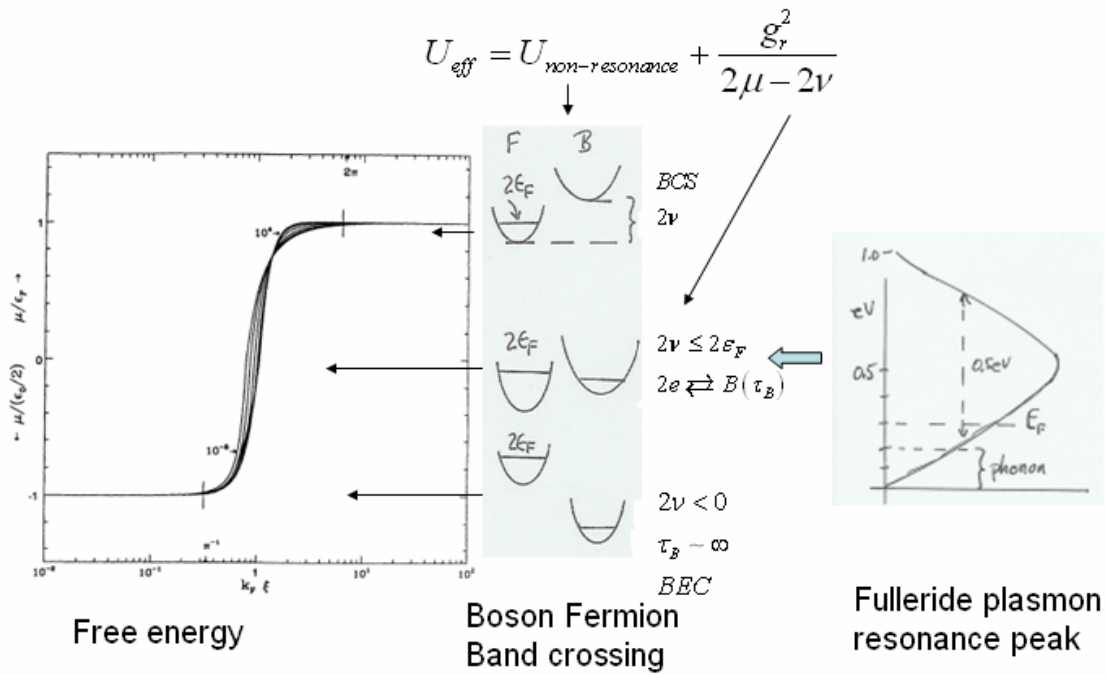

Fig 2. Correlation of crossover free energy [14] with the positions of the Fermion (F) and Boson (B) and the fulleride resonance peak. The crossover is quite literally a crossing of the Boson-Fermion bands from BCS to BEC with the result that the BCS pairing becomes dominated by the Feshbach resonance. In fullerides the BEC limit is not reached.